\documentclass{an}
\usepackage{graphicx}
\usepackage{times}
\usepackage{fancyhdr}
\begin{document}

\title{Magnetic fields in the Southern Galactic Plane Survey}

\author{Marijke Haverkorn \and
        Bryan M. Gaensler \and
        Jo-Anne C. Brown \and
        Naomi M. McClure-Griffiths \and
        John M. Dickey \and
        Anne J. Green}
\institute{Harvard-Smithsonian Center for Astrophysics; current
        address Astronomy Department, University of California-Berkeley \and
           Harvard-Smithsonian Center for Astrophysics \and
           University of Calgary \and
           Australia Telescope National Facility-CSIRO \and
           University of Tasmania \and
           University of Sydney}

\date{Received; accepted; published online}

\abstract{The Southern Galactic Plane Survey (SGPS) is a 1.4~GHz radio
  polarization and H~{\sc i} survey in a large part of the inner
  Galactic plane at a resolution of about an arcmin. Depolarization
  and Faraday rotation of polarized radiation from diffuse Galactic
  synchrotron emission, pulsars, and extragalactic sources can be used
  to infer information about the strength and structure of the
  Galactic magnetic field. Here, we discuss science results of the
  polarization data from the SGPS. We show from statistical analysis
  of rotation measures of polarized extragalactic sources that
  fluctuations in the magneto-ionized medium of the spiral arms are
  probably mainly caused by H~{\sc ii} regions, while the rotation
  measure fluctuations in the interarm regions may be connected to the
  interstellar turbulent cascade. Furthermore, the variations of
  rotation measure with Galactic longitude enable modeling of the
  large-scale component of the Galactic magnetic field, including
  determination of the number and location of magnetic field
  reversals. Finally, the SGPS is an excellent way to study
  subparsec-scale structure in the ionized ISM by way of
  depolarization studies in H~{\sc ii} regions. 
}

\correspondence{mhaverkorn@cfa.harvard.edu}

\maketitle

\section{Introduction}

Despite the pivotal role of the Milky Way magnetic field in many
physical processes such as turbulence in the interstellar medium
(ISM), heating of the gas, star formation and cosmic ray propagation,
details about its strength and structure remain elusive. This is in
large part because of the difficulty in measuring magnetism: the known
methods of observing the Galactic magnetic field are all indirect and
only detect a certain component of the field, or only in a certain
phase of the medium.

Magnetism in dense gas such as molecular clouds can be probed using
Zeeman splitting, whereas magnetic fields in dust are revealed by
optical polarization of starlight as well as submillimeter polarized
emission from the dust itself (e.g. Heiles 1987). A unique method to
explore magnetic fields in ionized gas in the Milky Way and external
galaxies is radio polarization. The total intensity is a measure of
the strength of the total magnetic field, whereas the polarized
intensity gives information about the regular magnetic field component
(e.g.\ Beck 2001). Faraday rotation and depolarization characteristics
are used to study the strength and structure of both regular and
random components of the magnetic field along the line of sight
(Gaensler et al.\ 2001; Brown \& Taylor 2001; Haverkorn, Katgert \& de
Bruyn\ 2004a, 2004b).

In this paper, we will describe the information about the Galactic
magnetic field obtained from the Southern Galactic Plane Survey. 
Therefore, in Section~\ref{s:sgps} we introduce the survey, and we
discuss some of the early science results in Section~\ref{s:science}.

\section{The Southern Galactic Plane Survey}
\label{s:sgps}

\begin{figure*}[t]
  \resizebox{\hsize}{!}
  {\includegraphics[]{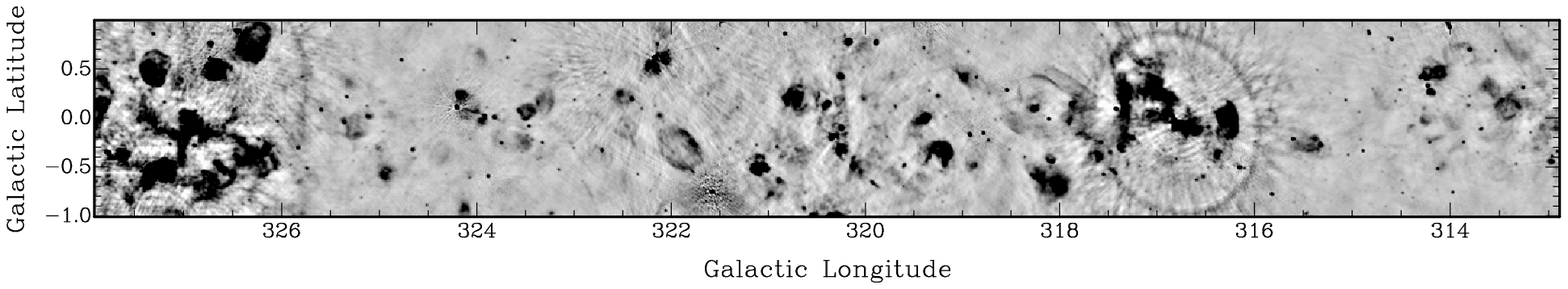}}
  \resizebox{\hsize}{!}
  {\includegraphics[]{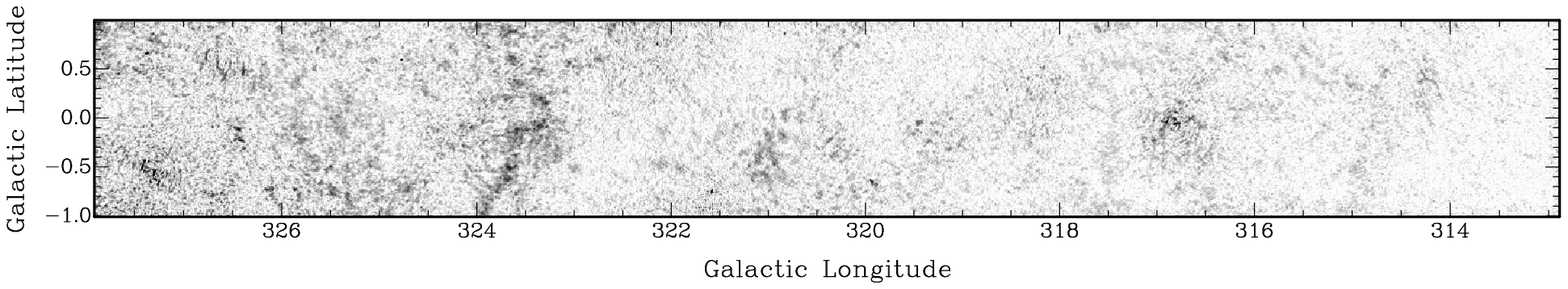}}
  \resizebox{\hsize}{!}
  {\includegraphics[]{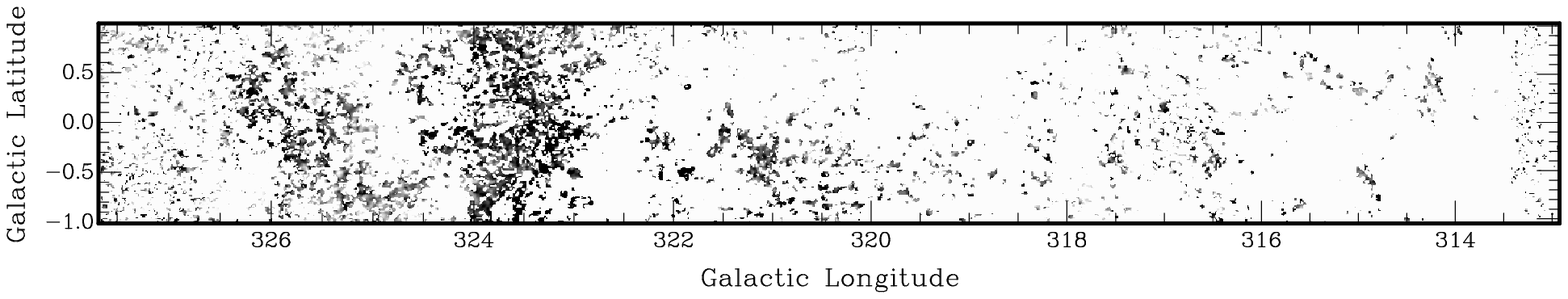}}
  \caption{Part of the Southern Galactic Plane Survey in total
           intensity $I$ at 1.4~GHz where max$(I) = 80$~mJy/bm (top),
           polarized intensity $P$ at 1.4~GHz where max$(P) =
           7$~mJy/bm (center), and rotation measure in the range $-100 <
           \mbox{RM} < 100$~rad~m$^{-2}$ (bottom). The large grating
           rings in $I$ (e.g.\ 
           around $l=317\degr$) are artifacts of the 15m regular
           spacing of the ATCA antennas. The bright sources that cause
           the grating rings create spurious polarization as well.} 
  \label{f:sgps}
\end{figure*}

The Southern Galactic Plane Survey (SGPS) is a survey in the neutral
hydrogen line and full-polarization continuum at 1.4~GHz, obtained
with the Australia Telescope Compact Array (ATCA) and Parkes 64m
single dish (McClure-Griffiths et al.\ 2005). (However, the data
presented here are solely from the ATCA.) The coverage of the
SGPS is $253\degr < l < 358\degr$ and $5\degr < l < 20\degr $ in
Galactic longitude, and $|b| < 1.5\degr$ in latitude, with a
resolution of $\sim1$~arcmin and a sensitivity of
$\sim1$~mJy/beam. The continuum data consist of 12 8~MHz-wide
full-polarization frequency bands from 1332~MHz to 1436~MHz. The
observables, Stokes parameters $I$, $Q$, $U$, and $V$, are translated
into polarization angle $\phi=0.5\arctan(U/Q)$ and debiased polarized
intensity $P=\sqrt{Q^2+U^2-\sigma^2}$, assuming that circular
polarization $V=0$.

The polarized radiation is Faraday rotated\footnote{Faraday  
  rotation of polarization angle $\phi$ occurs due to the
  birefringence of left and right circularly polarized radiation
  traveling through a medium which is ionized and magnetized. Faraday
  rotation is proportional to the wavelength of the radiation squared,
  i.e.\ $\Delta\phi = \mbox{RM} \lambda^2$. The rotation measure (RM)
  is given by RM~$\propto\int n_e {\mathbf B}\cdot {\mathbf {dl}}$,
  where $n_e$ is the electron density in the medium, ${\mathbf B}$ the
  magnetic field, and ${\mathbf {dl}}$ the path length through the
  medium.}  while propagating through the Galactic magneto-ionized
ISM, which allows determination of rotation measures (RMs). As an
example, a part of the SGPS is given in Fig.~\ref{f:sgps}, which shows
total intensity $I$, 
polarized intensity $P$, and RM. The polarized emission consists of
point sources (mostly extragalactic but also some pulsars) and diffuse
emission. The diffuse polarized and total intensities are mostly
uncorrelated, indicating Faraday rotation and depolarization. As can
be seen from Fig.~\ref{f:sgps}, RMs could only be determined for a
small part of the field. In the rest of the field, there is either no
polarized radiation due to complete depolarization, or
$\Delta\phi\neq\mbox{RM}\lambda^2$ due to missing short spacings
and/or partial depolarization.  

For a detailed description of the continuum part of the
SGPS discussed here see Haverkorn et al.\ (2006); the neutral
hydrogen part is described in McClure-Griffiths et al.\ (2005).

\section{Results}
\label{s:science}

\subsection{An additional source of RM fluctuations in the spiral arms
  only} 

\begin{figure}
  \resizebox{\hsize}{!}
  {\includegraphics[]{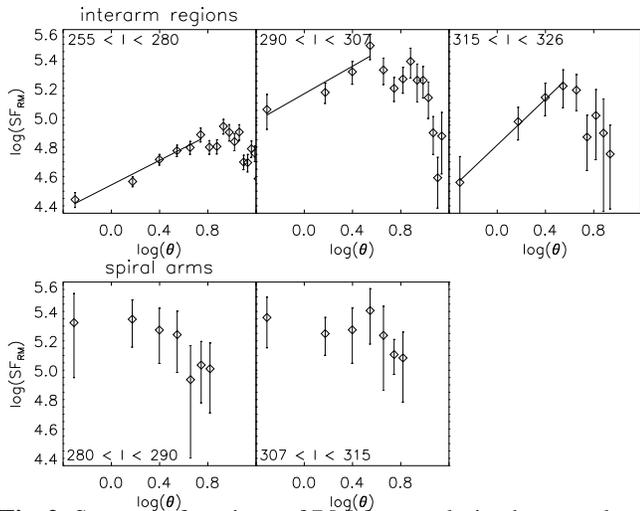}}
  \caption{Structure functions of RM from polarized extragalactic
           point sources as a function of angular scale $\theta$ for
           five regions in the SGPS. The top panels show interarm
           regions, the bottom panels spiral arms.}
  \label{f:sf}
\end{figure}

Statistical analysis of the RMs in the SGPS yields information about
the scale of RM fluctuations in the Galactic plane. The five panels in
Fig.~\ref{f:sf} each show the structure function\footnote{The (second
  order) structure function SF of a function $f$ as a function of
  separation $dx$ is defined as SF$_f(dx) =
  \langle(f(x)-f(x+dx))^2\rangle_x$, where $\langle\rangle_x$ denotes
  averaging over all positions $x$. Structure functions rise for
  increasingly larger fluctuations on increasingly larger scales $dx$,
  and saturate to a constant value at the maximum scale of structure
  present.} of RM in five regions in
the SGPS: the three top panels show interarm regions, whereas the
bottom two panels represent RM data in the Crux and Carina spiral
arms (Haverkorn et al.\ 2005). 

A clear difference in the behavior of the structure functions in
interarm regions and in spiral arms can be seen: the structure
functions in interarm regions are all rising, whereas those in the
spiral arms are consistent with flat. 
The rising structure functions in the interarm regions denote
correlated RM fluctuations in the interarm regions up to the angular
scales where the structure function turns over, i.e.\ several
degrees. The power law in the structure function is reminiscent of
turbulence, as is observed in velocity spectra, and the observed
slopes agree with velocity spectra slopes for incompressible
(magneto-)hydrodynamical turbulence (Kolmogorov 1941; Goldreich \&
Sridhar 1995). This suggests coupling of the observed density spectrum
to a Kolmogorov-like velocity spectrum, which can only exist if the
turbulence in the ionized ISM is nearly incompressible and not highly
supersonic. Indeed, non-thermal linewidths in Galactic H$\alpha$ seem
to indicate transonic or only mildly supersonic turbulence (Reynolds
1985; Tufte, Reynolds \& Haffner 1999). 

On the other hand, {\it in the spiral arms} the structure functions
are flat, i.e.\ the scale at which the structure function saturates is
smaller than the scales sampled in these observations. As
the RM is an integral along the whole line of sight, it is not
possible to attribute a certain distance to the medium where the RM
originates. However, assuming that the largest-scale structure comes
from the nearby ISM, we can estimate an upper limit for the outer
scale of structure in the Carina arm of about 17~pc, assuming that the
nearest gas from that arm is at $\sim2$~kpc (Russeil 2003). An obvious
source for these fluctuations would be internal structure in H~{\sc
  ii} regions, which are sufficiently abundant in the spiral arms and
of roughly the right size (Haverkorn et al.\ 2004c).

\subsection{Depolarization around H~{\sc ii} regions}
\label{ss:rcw}

\begin{figure}
  \begin{center}
    {\includegraphics[angle=-90,width=0.86\hsize]{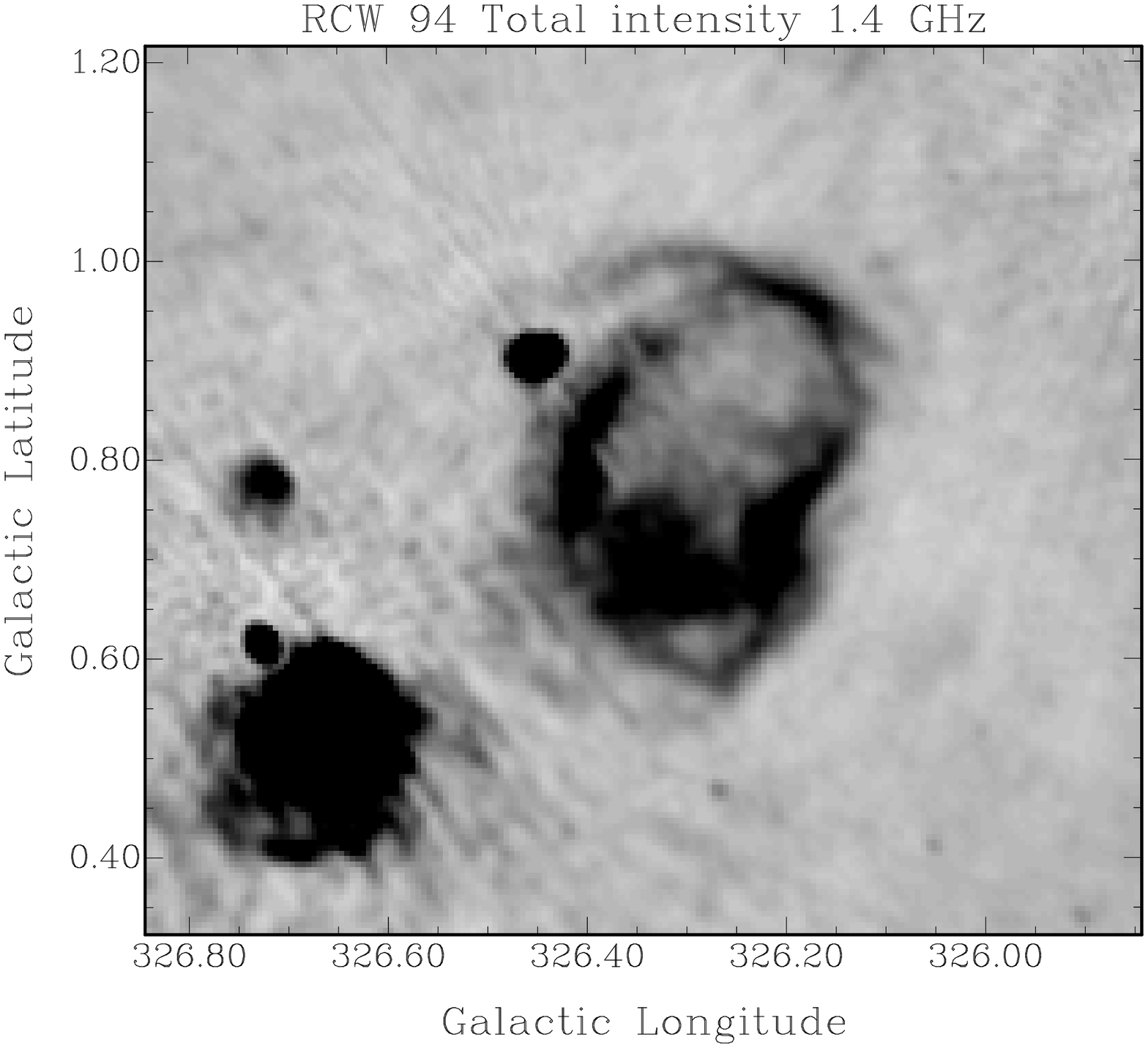}}
    {\includegraphics[angle=-90,width=0.86\hsize]{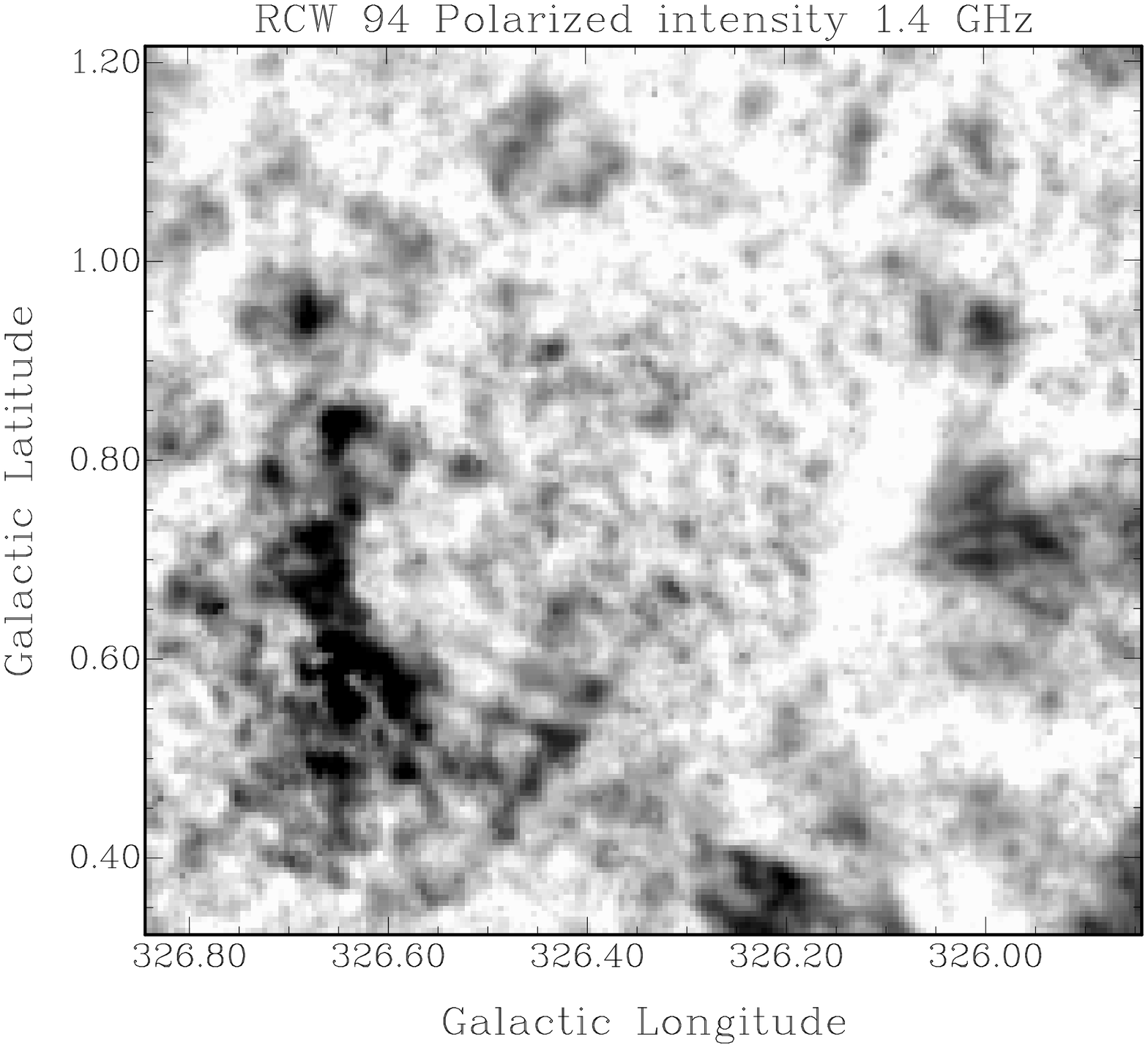}}
  \end{center}
  \caption{Total (top) and polarized (bottom) intensity in the H~{\sc
           ii} region RCW~94, showing depolarization at the location
           of RCW~94 which is strongest just outside the radio
           continuum edge, indicating a depolarization halo.}
  \label{f:rcw94}
\end{figure}

H~{\sc ii} regions do not emit polarized radiation. However, polarized
radiation coming from behind an H~{\sc ii} region can be depolarized
by it if strong enough fluctuations in the plasma density and/or
magnetic field are present on scales smaller than the synthesized
beam. From the amount of depolarization by an H~{\sc ii} region,
combined with an estimate of its electron density, the strength of its
random magnetic field and the outer scale of the fluctuations can be
estimated. A beautiful example in the SGPS Test Region is shown in
Fig.~\ref{f:rcw94}. The bottom map shows a depolarized shell around
the H~{\sc ii} region, which extends beyond the radio continuum
emission shown in the top map. This is evidence for a depolarization
halo around the H~{\sc ii} region, probably caused by interaction with
a molecular cloud (Gaensler et al.\ 2001). The H~{\sc ii} region
dissociates H$_2$ molecules from the molecular cloud, which results in
a shell of H~{\sc i} around the H~{\sc ii} region, as observed
(McClure-Griffiths et al.\ 2001).

A depolarization signature like the one observed cannot be produced by
a spherical H~{\sc ii} region. Instead, the RM must be almost constant
in the central part of the region, which is in agreement with H$\alpha$
observations (Georgelin et al.\ 1994). The almost constant degree of
polarization in the interior of the H~{\sc ii} region can be modeled
with a random magnetic field of $\sim 1.2~\mu$G, varying on scales of
approximately 0.2~parsec (Gaensler et al.\ 2001).

\subsection{Probing the regular Galactic magnetic field}

As opposed to the diffuse polarized emission discussed in
Section~\ref{ss:rcw}, RMs of unresolved polarized extragalactic
sources do not suffer from depolarization. Therefore, RMs of polarized
extragalactic sources represent the complete line of sight through the
Milky Way. This, combined with their abundance in the SGPS, makes them
ideal probes of the regular component of the Galactic magnetic field. 

\begin{figure}
  \resizebox{\hsize}{!}
  {\includegraphics[]{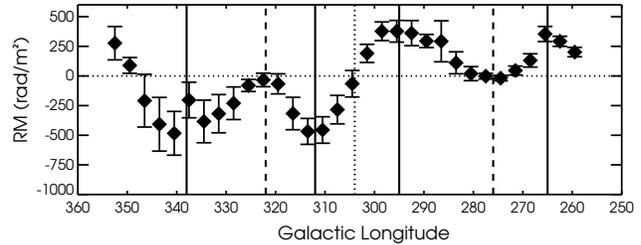}}
  \caption{RM as a function of Galactic longitude. The RM is averaged
           over 9\degr\ in Galactic longitude and the spacing between
           the data points is 3\degr, so that the points are not
           independent. The solid lines are RM maxima, the dashed
           lines are RM minima, and the dotted line denotes a
           large-scale sign change in RM.}
  \label{f:rml}
\end{figure}

Fig.~\ref{f:rml} shows the distribution of RMs of extragalactic point
sources as a function of Galactic longitude, binned in 9\degr\ bins
with 3\degr\ separation to average out the small-scale structure and 
the intrinsic RM of the sources (RM$_{int} \la 5$~rad~m$^{-2}$). From
the Figure, smooth large-scale structure on scales of tens of
degrees is evident, believed to be caused by the magnetic field
structure along the spiral arms. E.g.\ at longitude
$l\approx304\degr$, the smoothed RM changes sign, which means that the
magnetic field averaged over the line of sight changes direction. A
regular sign change in RM like this can only happen if there is a
large-scale reversal of the magnetic field direction at this
longitude. This reversal near the Carina arm has been noted many times
before (see e.g.\ Vall\'ee 2002).

\begin{figure}
  \resizebox{.95\hsize}{!}
  {\includegraphics[]{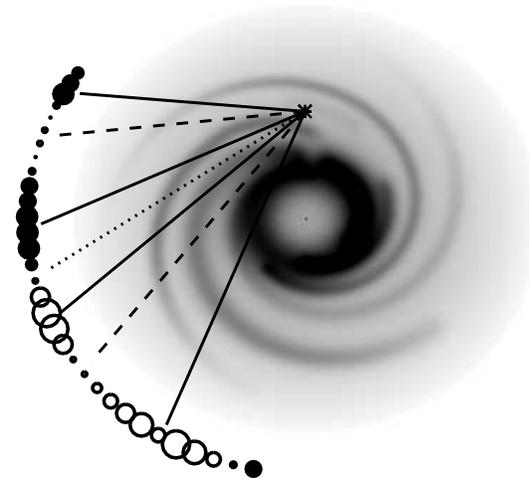}}
  \caption{Bird's eye view of the Galaxy, where the gray scale denotes
           the NE2001 electron density model and circles represent RMs
           of extragalactic sources as binned in
           Fig.~\ref{f:rml}. Open (closed) circles are negative
           (positive) RMs, and the largest circle denotes
           RM$=-484$~rad~m$^{-2}$. The lines originating in the
           position of the Sun are the same lines of sight as in
           Fig.~\ref{f:rml}.}
  \label{f:cl}
\end{figure}

The lines in Fig.~\ref{f:rml} mark positions where the RM is extreme
or passes through zero. The directions of these lines of sight are
given as well in Fig.~\ref{f:cl}, which is a bird's eye view of the
Milky Way. The gray scale denotes the electron density model NE2001
(Cordes \& Lazio 2003) and the circles represent binned RMs of
extragalactic sources set at an arbitrary distance outside the
Galaxy. From Fig.~\ref{f:cl} it can be seen that $|$RM$|$ is roughly
maximum at sight lines through spiral arms in the NE2001 model, and
minimum in interarm regions. A smooth decrease of $|$RM$|$ to almost
zero can only occur if there are large-scale magnetic field reversals
along the line of sight, although not necessarily on Galactic scales.
Fitting the RM profile to models of the regular magnetic
field yields information about the strength of the regular
magnetic field component, and about the number and location of
reversals (Brown et al. 2006).

Note that although reversals in the regular magnetic field direction
must be on scales of tens of degrees to cause these large-scale RM
variations, the sampled Galactic longitude range is too small to be
able to say if the reversal or reversals trace the spiral arms around
the entire Galaxy. Galaxy-scale reversals in magnetic field that
follow the spiral arms seem to fit the present data, however,
only very few reversals have been observed in external galaxies (Beck
2001). 

\acknowledgements 

The ATCA is part of the Australia Telescope, which is funded by the
Commonwealth of Australia for operation as a National Facility managed
by CSIRO. MH and BMG acknowledge support from the National Science
Foundation through grant AST-0307358 to Harvard University.


\begin{thebibliography}{}
\bibitem[Beck(2001)]{b01} 
  Beck, R. 2001: SSRv~99, 243
\bibitem[Brown \& Taylor(2001)]{bt01} 
  Brown, J. C., \& Taylor, A. R. 2001, ApJ, 563, L31
\bibitem[Brown et al.(2006)]{b06}
  Brown, J. C., et al.: 2006, in prep
\bibitem[Cordes \& Lazio(2003)]{cl03} 
  Cordes, J.~M., Lazio, T.~J.~W.: 2003, preprint (astro-ph/0301598)
\bibitem[Gaensler et al.(2001)]{gdm01} 
  Gaensler, B.~M., Dickey, J.~M., McClure-Griffiths, N.~M., Green,
  A.~J., Wieringa, M.~H., Haynes, R.~F.: 2001, ApJ~549, 959
\bibitem[Georgelin et al.(1994)]{gag04}
  Georgelin, Y. M., Amram, P., Georgelin, Y. P., Le Coarer, E.,
  Marcelin, M.: 1994, A\&AS~108, 513
\bibitem[Goldreich \& Sridhar(1995)]{gs95} 
  Goldreich, P., Sridhar, S.: 1995, ApJ~438, 763 
\bibitem[Haverkorn et al.(2004a)]{hkb04a} 
  Haverkorn, M., Katgert, P., de Bruyn, A. G.: 2004a, A\&A~427, 169  
\bibitem[Haverkorn et al.(2004b)]{hkb04b} 
  Haverkorn, M., Katgert, P., de Bruyn, A. G.: 2004b, A\&A~427, 549
\bibitem[Haverkorn et al.(2004c)]{hgm04c}
  Haverkorn, M., Gaensler, B.~M., McClure-Griffiths, N.~M., Dickey,
  J.~M., Green, A. J.: 2004c, ApJ~609, 776 
\bibitem[Haverkorn et al.(2005)]{hgb05}
  Haverkorn, M., Gaensler, B.~M., Brown, J. C., Bizunok, N. S.,
  McClure-Griffiths, N.~M., Dickey, J.~M., Green, A. J.: 2005,
  ApJ~submitted 
\bibitem[Haverkorn et al.(2006)]{hgm06}
  Haverkorn, M., Gaensler, B.~M., McClure-Griffiths, N.~M., Dickey,
  J.~M., Green, A. J.: 2006, in prep
\bibitem[Heiles(1987)]{h87} 
  Heiles, C.: 1987, in {\it Interstellar processes}. Dordrecht,
  D. Reidel Publishing Co.
\bibitem[Kolmogorov(1941)]{k41} 
  Kolmogorov, A.~N.: 1941, Dokl.\ Akad.\ Nauk SSSR~30, 301
\bibitem[McClure-Griffiths et al.(2001)]{mdg01} 
  McClure-Griffiths, N. M., Dickey, J. M., Gaensler, B. M., Green,
  A. J., Haynes, R. F., Wieringa, M. H.: 2001, PASA~18, 84
\bibitem[McClure-Griffith et al.(2005)]{mdg05}
  McClure-Griffiths, N.~M., Dickey, J.~M., Gaensler, B.~M., Green,
  A.~J., Haverkorn, M., Strasser, S.: 2005, ApJS~158, 178
\bibitem[Reynolds(1985)]{r85}
  Reynolds, R. J.: 1985, ApJ~294, 256
\bibitem[Russeil(2003)]{r03}
  Russeil, D.: 2003, A\&A~397, 133
\bibitem[Taylor et al.(2003)]{tgp03}
  Taylor, A.~R., Gibson, S.~J., Peracaula, et al.: 2003, AJ~125, 3145 
\bibitem[Tufte et al.(1999)]{trh99}
  Tufte, S., Reynolds, R., Haffner, M.: 1999, in {\it Interstellar
  Turbulence}, eds.\ J.\ Franco and A.\ Carrami\~nana, Cambridge
  University Press 
\bibitem[Vall\'ee(2002)]{v02} 
  Vall\'ee, J. P.: 2002, ApJ~566, 261
\end{thebibliography}
\end{document}